\documentclass[12pt]{article}

\usepackage{amssymb}
\usepackage{amsmath}
\usepackage{graphics}
\usepackage{latexsym}
\usepackage{array}
\usepackage[activeacute,english]{babel}
\usepackage[latin1]{inputenc}
\usepackage[dvips]{graphicx}
\usepackage{color}

\newcommand{\bn}{\begin{eqnarray}}
\newcommand{\en}{\end{eqnarray}}
\newcommand{\bq}{\begin{equation}}
\newcommand{\eq}{\end{equation}}

\newcommand{\bc}{\begin{center}}
\newcommand{\ec}{\end{center}}

\title{\large \bf SU(3) Maxwell equations and the classical chromodynamics}
\author{\large J. A. Sánchez-Monroy{\footnote{E-mail address:
jasanchezm@unal.edu.co.}} and C. J. Quimbay{\footnote{E-mail
address: cjquimbayh@unal.edu.co.  Associate researcher of the Centro
Internacional de F\'{\i}sica, Ciudad Universitaria, Bogot\'{a} D.
C., Colombia.}}}
\date{{\it Departamento de F\'{\i}sica, Universidad Nacional de Colombia.\\
Ciudad Universitaria, Bogot\'{a} D.C., Colombia.}\\}
\begin{document}
\maketitle

\begin{center} {\bf \large Abstract} \end{center}

\noindent {\footnotesize We study the equations of motion of the
SU(3) Yang-Mills theory. Since the gluons, at scales of the order of
$1$ $fm$, can be considered as classical fields, we suppose that the
gauge fields ($A_\mu^a$) of this theory are the gluonic fields and
then it is possible to consider the Quantum Chromodynamics in a
classical regime. For the case in which the condition $[A_\mu^a,
A_\rho^a]=0$ is satisfied, we show that the abelian equations of
motion of the Classical Chromodynamics (CCD) have the same form as
those of the classical electrodynamics without sources.
Additionally, we obtain the non-abelian Maxwell equations for the
CCD with sources. We observe that there exist electric and magnetic
colour fields whose origin is not fermionic. We show as the gluons
can be assumed as the sources of the electric and magnetic colour
fields. We note that the gluons are the only responsible for
the existence of a magnetic colour monopole in the CCD.\\

}

\section{Introduction}\label{intro}

Hadrons are composed by quarks which interact among them through the
in\-ter\-me\-dia\-te boson fields of the strong interaction
(gluons). The strong interaction between quarks and gluons is
described by the Quantum Chromodynamics (QCD), a non-abelian gauge
theory based in the $SU(3)_c$ colour gauge symmetry. The QCD can
describe the observed phenomenology in the high-energy regime
through the use of perturbation theory. However, for the low-energy
regime, it is necessary to use non-perturbative methods because the
running coupling constant is large. In this non-perturbative regime,
in which the quarks are confined forming hadrons, several methods
have been developed, as for instance, lattice theory, relativistic
quarks mo\-dels and effective theories. In the context of a
particular class of relativistic quarks model \cite{yu1} is possible
to describe the charmonium and bottomonium spectrum, in good
agreement with the experimental data, by solving the Dirac equation
in pre\-sen\-ce of SU(3) Yang-Mills fields representing gluonic
fields. The obtaining solutions can model the quark confinement in a
satisfactory way. At this respect, the results shown in \cite{yu1}
suggest that the mechanism of quark confinement should occur within
the framework of QCD. Explicit calculations performed by Yu
Goncharov \cite{yu2, yu3} show how the gluon concentration is huge
at scales of the order of $1$ fm. This fact suggests that the
gluonic fields form a boson condensate and therefore, the gluons at
large distances can be considered as classical fields \cite{yu4}.

The main goal of this paper is to study the equations of motion of
the SU(3) Yang-Mills theory. For this theory, we suppose that the
classical gauge fields ($A_\mu^a$) represent the gluonic fields. At
large distances, of the order of $1$ fm, these fields are assumed as
classical fields and then the theory is a classical version of the
QCD. For the case in which the condition $[A_\mu^a, A_\rho^a]=0$ is
satisfied, we show that the abelian equations of motion of the
Classical Chromodynamics (CCD) have the same form as those of the
classical electrodynamics without sources. Additionally, if we
consider the sources as having only one colour charge and assume
that there only exist two diagonal gluon fields, we find that the
equations of motion of the CCD also have the same form as those of
the electrodynamics with sources. On the other hand, for the $SU(3)$
Yang-Mills theory with sources, we obtain the non-abelian Maxwell
equations. We observe that the divergence of the electric and
magnetic colour fields are non vanishing. The latter implies that
there exist electric and magnetic colour sources whose origin is not
fermionic. The origin of these sources is related to the fact that
the gluons have colour charge and therefore they can be assumed as
the sources of the electric and magnetic colour fields. We note that
the gluons are the only responsible for the existence of a magnetic
colour monopole in the CCD.\

\section{Equations of motion of the classical chromodynamics}

The equations of motion of the SU(3) Yang-Mills theory with quarks
sources $J_b^\nu$ are:

\begin{equation}\label{yms}
\partial _\mu F_b ^{\mu
\nu}+gC^c _{ab}A^a _\mu F_c ^{\mu \nu}=gJ_b^\nu=g\bar \psi \gamma
^\nu \lambda_b\psi,
\end{equation}
being $F_{\mu \nu}^b$ the non-abelian gauge field tensor,
$\lambda_b$ the Gell-Mann matrices, $g$ the running coupling
constant of the $SU(3)$ group, $C^c _{ab}$ the structure constants
of the Lie algebra associated to this gauge group and $\psi$ the
quark field. The equations of motion given by (\ref{yms}) represents
a system of non-lineal equations, whose solutions are supposed to
contain at least the components of the SU(3) field which are Coulomb
like or linear in the distance between quarks $(r)$. In this way,
the solutions of (\ref{yms}) can model the quark confinement
\cite{yu1}. Since the Coulomb potential is solution of the equations
of motion of the classic electrodynamics in a problem with fermionic
sources, it is relevant to ask us under which conditions the motion
equations of the chromodynamics have the same form of those of the
electrodynamics. The main goal of this section is to answer this
question starting from the SU(3) Yang-Mills motion equations given
by (\ref{yms}).

Since the Lie algebra of the $SU(3)$ gauge group is defined by
$[\lambda _a,\lambda _b]=C_{ab}^c\lambda ^c$ and rewriting the
non-abelian gauge fields as $A_\mu =A_\mu ^a \lambda_a$, it is
possible to write the non-abelian gauge field tensor as:
\begin{equation}\label{ym2}
F_{\mu \nu}=\partial _\mu A_\nu -\partial _\nu A_\mu -ig[A_\mu,
A_\rho].
\end{equation}
In the last expression, the colour index does not appear explicitly
and each term is a matrix. For the case in which the quark sources
vanish, i. e. for $J_b^\nu=0$, the SU(3) Yang-Mills equations of
motion given by (\ref{yms}) can be written as
\begin{equation}\label{tgf}
\partial ^\mu F_{\mu \nu}=-ig[A^\mu, F_{\mu \nu}].
\end{equation}
Substituting (\ref{ym2}) in (\ref{tgf}), the equations of motion of
the CCD without sources are
\begin{equation}\label{ccdme1}
\partial ^\mu (\partial _\mu A_\nu -\partial _\nu A_\mu -ig[A_\mu, A_\rho])
=-ig[A^\mu, (\partial _\mu A_\nu -\partial _\nu A_\mu -ig[A_\mu,
A_\rho])] ig[A_\mu, [A_\mu, A_\rho]]).
\end{equation}

We observe that the system of equations given by (\ref{ccdme1}) is
clearly not lineal, implying that the solutions cannot be obtained
easily. A special case of (\ref{ccdme1}) corresponds to the
situation in which the abelian condition given by
\begin{equation}\label{abelcon}
[A_\mu, A_\rho]=0
\end{equation}
is satisfied. This condition, i.e $[A_\mu ^a \lambda_a, A_\nu ^b
\lambda_b]=0$, can be satisfied in a non-trivial way if and only if
one of the two following conditions is satisfied: ${\bf \it i}$) If
only the Gell-Mann matrices of the Cartan subalgebra, which is a
maximal abelian of the $SU(3)$ gauge group Lie algebra, appear in
the system of equations of motion; ${\bf \it ii}$) If $A_\mu ^a
\lambda_a=mA_\nu ^a \lambda_a$, being $m$ a constant. The first
condition means that there only exist two gluon fields in the
system, precisely the associated with the $\lambda_3$ and
$\lambda_8$ generators. These generators are shown in Appendix A.
The second condition implies that each component of $A_\mu$ is
transmitted through the same gluonic configuration. This last
possibility is not clear from a physical point of view.

Applying the condition (\ref{abelcon}) in (\ref{ccdme1}), we obtain
that the abelian equations of motion of the SU(3)Yang-Mills theory
are
\begin{equation}\label{qed}
\partial ^\mu F'_{\mu \nu}=0,
\end{equation}
being $F'_{\mu \nu}=\partial _\mu A_\nu -\partial _\nu A_\mu$, and
$A_\mu$ given by
\begin{equation}\label{mat}
A_{\mu}^a\lambda_a = \left(
\begin{array}{ccc}
A_{\mu}^3+\frac{1}{\sqrt{3}}A_{\mu}^8 & A_{\mu}^1-iA_{\mu}^2 & A_{\mu}^4-iA_{\mu}^5\\
A_{\mu}^1+iA_{\mu}^2 & -A_{\mu}^3+\frac{1}{\sqrt{3}}A_{\mu}^8 & A_{\mu}^6-iA_{\mu}^7\\
A_{\mu}^4+iA_{\mu}^5 & A_{\mu}^6+iA_{\mu}^7 & -\frac{2}{\sqrt{3}}A_{\mu}^8
\end{array}
\right).
\end{equation}
We observe that there exist two differential equations for each one
of the first seven fields and three for the eighth field. It is very
easy to probe that the abelian equations of motion (\ref{qed}) can
be written as:
\begin{equation}\label{qed2}
\partial ^\mu F'^a_{\mu \nu}=0,
\end{equation}
which means that there exists an equation similar to (\ref{qed}) for
each field $A^a_{\mu}$. We note that the equations (\ref{qed2}) have
the same form as the electrodynamics ones. Under this similarity,
the equations of motion of the CCD have the same behaviour as those
of the classical electrodynamics. All the solutions of the equations
(\ref{qed}) are also solutions of the equations (\ref{yms}) with
$J_b^\nu=0$.

The Yang-Mills equations given by (\ref{qed2}) can be written in a
similar form as the Maxwell equations of the electrodynamics without
sources. These abelian Maxwell equations for the CCD are:
\begin{eqnarray}\label{qen}
\vec{\nabla} \cdot \vec{E}_c&=&0,\\
\vec{\nabla} \times \vec{E}_c&=&-\frac{\partial \vec{B}_c}{\partial t},\\
\vec{\nabla} \cdot \vec{B}_c&=&0,\\
\vec{\nabla} \times \vec{B}_c&=&\frac{1}{c^2} \frac{\partial
\vec{E}_c}{\partial t},\label{qen4}
\end{eqnarray}
being $\vec{E}_c$ and $\vec{B}_c$ the electric and magnetic colour
fields, respectively. Using the equations (\ref{qen})-(\ref{qen4})
is possible to predict the existence of CCD waves.

Now, we consider the equations of motion of the SU(3) Yang-Mills
theory with sources, i. e. for the case $J\not=0$. Following a
procedure similar as the $J=0$ case and demanding that the equations
of motion satisfy the abelian condition $[A_\mu, A_\rho]=0$, we can
write the equations of motion (\ref{yms}) as
\begin{equation} \label{qee}
\partial _\mu F'^{\mu \nu}=\lambda^a\bar \psi \gamma ^\nu
\lambda_a\psi.
\end{equation}
Because the quark field $\psi$ is a triplet in the colour space
\begin{equation}\label{psi}
\psi=\left( \begin{array}{c}\psi_{Red} \\ \psi_{Blue} \\
\psi_{Green} \end{array} \right)=\left( \begin{array}{c}\psi_{R}
\\ \psi_{B} \\ \psi_{G} \end{array} \right) ,
\end{equation}
then, the right side of (\ref{qee}) has the following form:
\begin{equation}
\footnotesize{ \left(
\begin{array}{ccc}
\frac{2}{3}(2\bar \psi_B  \psi_B-\bar \psi_R  \psi_R- \bar \psi_G
\psi_G) & 2\bar \psi_B \psi_R & 2\bar \psi_B \psi_G\\ 2\bar \psi_R
\psi_B & \frac{-2}{3}(\bar \psi_B  \psi_B-2\bar \psi_R  \psi_R+
\bar \psi_G  \psi_G) & 2\bar \psi_R  \psi_G \\2\bar \psi_G \psi_B
& 2\bar \psi_G  \psi_R & \frac{-2}{3}(\bar \psi_B \psi_B+ \bar
\psi_R \psi_R-2 \bar \psi_G  \psi_G)
\end{array}
\right) },
\end{equation}
where each element of the matrix has a $\gamma ^\nu$ between the
fields $\bar \psi \psi$. Since the left side of (\ref{qee}) is a
system of equations, as is shown in (\ref{mat}), it is possible to
consider the different situations in which the equation (\ref{qee})
is satisfied. Our interest is focused for the case in which the
Yang-Mills equations is uncoupled respect to the components in the
colour space of the quark fields. This case is considered if the
abelian condition $[A_\mu, A_\rho]=0$ is satisfied. For this abelian
case, there are only two gluonic fields, the associated with the
generators $\lambda_3$ and $\lambda_8$, different to zero and then
it is necessary that two components in the colour space of the quark
fields vanish independently. This means that for the CCD there only
exists one colour charge, in a similar way as what happens for the
electrodynamics in which there only exists one electric charge. This
analysis leads to assume that any solution for the electrodynamics
with sources is also a solution of the CCD. For instance, the
Coulomb potential:
\begin{eqnarray}
A_t=\sum ^N_{i=1}\frac{\alpha}{|\vec{r}-\vec{a}_i|},
\end{eqnarray}
is solution of the SU(3)-Yang-Mills equations of motion given by
(\ref{qed}). The latter assures that each fermionic source having
colour charge contributes to the Coloumb potential. It is possible
that there exist more potentials \cite{mate}, but this result only
assures that there exist at least one having the form of a Coloumb
potential. This result is important because is a justification to
take a Coloumb potential in the description of problems with three
or more quarks.


\section{Non-abelian Maxwell equations}
The electric ($\vec E$) and magnetic ($\vec B$) fields, in the
classical electrodynamics, are defined as the components of the
electromagnetic field tensor ($F_{\mu \nu}$), in the following form:
\begin{eqnarray}
E_i&:=&F_{0 i},\\
B_n&:=&-\frac{1}{2}\varepsilon _{nij}F_{ij},
\end{eqnarray}
being $n,i,j=1,2,3$. Starting from the Lagrangian of the
electromagnetic field with sources is possible to obtain the
Yang-Mills equations using the Euler-Lagrange equations. From these
equations is possible to obtain the homogeneous Maxwell equations.

In an analogous way, we consider the non-abelian Maxwell equations
for the CCD. For this case, the gauge field tensor is
\begin{eqnarray}\label{ym1}
F^a_{\mu \nu}&=&\partial _\mu A^a _\nu -\partial _\nu A^a_\mu +
gC^a_{bc}A^b_\mu A^c_\nu,
\end{eqnarray}
where the non-abelian gauge field is $A^a_\mu=(A^0, -\vec A)$. The
electric and magnetic colour fields, for the CCD, are defined
respectively as:
\begin{eqnarray}
E^a_i&:=&F^a_{0 i}=-\partial _0 A^a _i -\partial _i A^a_0 -
gC^a_{bc}A^b_0 A^c_i,\label{ecf}\\
B^a_j&:=&-\frac{1}{2}\varepsilon _{jik}F^a_{ik}=-\frac{1}{2}
\varepsilon _{jik}\left(-\partial _i A^a _k +\partial _k A^a_i
+gC^a_{bc}A^b_i A^c_k\right)\label{mcf},
\end{eqnarray}
being $n,i,j=1,2,3$. In vectorial notation, the electric and
magnetic colour fields are
\begin{eqnarray}
\vec{E}^a&=&-\partial_t \vec{A}^a-\vec{\nabla} A_0^a-gC^a_{bc}
A_0^b\vec{A}^c,\\
\vec{B}^a&=&\vec{\nabla} \times \vec{A}^a-\frac{1}{2}gC^a_{bc}
\left(\vec{A}^b \times \vec{A}^c\right).
\end{eqnarray}
In contrast with the magnetic field of the electrodynamics, the
magnetic colour field for the CCD is written as the sum of a rotor
term and a non-rotor term.

Using the SU(3) Yang-Mills equations of motion (\ref{yms}), we
obtain that the first Maxwell equation for the CCD with sources is
given by:
\begin{equation}
\partial^i E^a_{i}=-gC^a_{bc}A^b_i E^c_{i}+\rho^a,
\end{equation}
or in vectorial notation:
\begin{equation}\label{me1}
\vec{\nabla} \cdot \vec{E}^a=-gC^a_{bc}\vec{A}^b \cdot
\vec{E}^c+\rho^a,
\end{equation}
where we have used the fact that the fermionic source can be
written as $J^a_\mu=(\rho^a, -\vec J^a)$. Using the following
relation:
\begin{eqnarray}
B^a_j&:=&-\frac{1}{2}\varepsilon _{jik}F^a_{ik},\notag\\
\varepsilon _{jpq}B^a_j&:=&-\frac{1}{2}\varepsilon _{jpq}
\varepsilon _{jik}F^a_{ik},\notag\\
\varepsilon _{jpq}B^a_j&:=&-\frac{1}{2}(\delta_{pi}
\delta_{qk}-\delta_{pk}\delta_{iq})F^a_{ik},\notag\\
\varepsilon _{jpq}B^a_j&:=&-F^a_{pq},
\end{eqnarray}
then, it is possible to obtain the second Maxwell equation:
\begin{eqnarray}
\partial^\mu F^a_{\mu j}&=&-gC^a_{bc}A^b_\mu F^c_{\mu j}-J^a_j,\notag\\
\partial^0 E^a_{j}-\partial^i F^a_{i j}&=&-gC^a_{bc}A^b_0 E^c_{j}
+gC^a_{bc}A^b_i F^c_{i j}-J^a_j,\notag\\
\partial^0 E^a_{j}+\partial^i \varepsilon _{lij}B^a_l&=&-gC^a_{bc}A^b_0
E^c_{j}-gC^a_{bc}A^b_i \varepsilon _{lij}B^c_l-J^a_j,
\end{eqnarray}
that in vectorial notation can be written as
\begin{equation}\label{me2}
\vec{\nabla} \times
\vec{B}^a-\partial_t\vec{E}^a=\vec{J}^a+gC^a_{bc}A^b_0
\vec{E}^c-gC^a_{bc}\vec{A}^b\times \vec{B}^c.
\end{equation}
The other two Maxwell equations are obtained from the definitions of
the electric (\ref{ecf}) and magnetic (\ref{mcf}) colour fields:
 \begin{eqnarray}\label{me3}
\vec{\nabla} \cdot \vec{B}^a&=&-\frac{1}{2}gC^a_{bc}\nabla \cdot
{(\vec{A}^b \times \vec{A}^c)}
\end{eqnarray}
and
 \begin{eqnarray}\label{me4}
\vec{\nabla} \times
\vec{E}^a+\partial_t\vec{B}^a&=&\frac{1}{2}gC^a_{bc}
\partial_t\left(\vec{A}^b \times \vec{A}^c\right)-gC^a_{bc}
\left[\vec{\nabla} \times (A^b_0 \vec{A}^c)\right].
\end{eqnarray}

The Maxwell equations (\ref{me1}), (\ref{me2}), (\ref{me3}) and
(\ref{me4}) can be extended directly for any $SU(N)$ gauge group.
The interest here is for the $SU(3)$ gauge group. The SU(3)
Yang-Mills equations have solutions and these solutions are unique
for the case in which $\vec{E}$ and $\vec{A}$ are independent fields
and if there exist particular boundary conditions \cite{mate}.

The wave equations for the CCD are given by (see Appendix B):
\begin{eqnarray}
\Delta A^a _\nu&=&gC^a _{bc}A^b _\mu (\partial _\nu A^c_\mu-
2\partial _\mu A^c _\nu-gC^c_{mn}A^m_\mu A^n_\nu).
\end{eqnarray}

\section{Conclusions}

In the first part of this paper, we have found that the abelian
equations of motion for the CCD without sources have the same form
as those of electrodynamics without sources. These equations of
motion have been obtained imposing that the SU(3) Yang-Mills
equations of motion satisfying the abelian condition given by
$[A_\mu, A_\rho]=0$. Starting of this result is possible to predict
the existence of waves for the CCD. When considering sources, using
the same abelian condition, it is possible to find Yang-Mills
equations uncoupled to the components in the colour space of the
quark fields. We have found, under the abelian condition, that there
only exist one colour charge, reminding us the electrodynamics case
where there only exists one electric charge. For this case, as
maximum there are two gluonic fields as the mediators of the strong
interaction. The equations of motion of the CCD with sources have
the same form as those of the electrodynamics with sources, for each
one of the two independent bosons. This last result assures that in
a system with $N$ quarks, each quark contributes with a Coulomb
potential. Surely in this problem there exist more potentials, but
at least one of them is a Coulomb potential.

In the second part of the paper, we have obtained the equations of
motion for the CCD analogous to the Maxwell equations for the
electrodynamics. These non-abelian Maxwell equations have been
obtained for a SU(3) Yang-Mills theory, but they are directly
extended for a SU(N) Yang-Mills theory. We have found that the
Maxwell equations does not only depend on $\vec{E}^a$ and
$\vec{B}^a$ but also on $\vec{A}^a$ and $A^a_0$. From the
divergences of $\vec{E}^a$ and $\vec{B}^a$, it is possible to
conclude that there exist sources of electric and magnetic colour
fields which are not fermions. It is possible to see as well that
the bosonic field is charged and simultaneously is source of
magnetic field, i. e. the gluonic fields have colour charge.
Additionally, as the divergence of $\vec{B}^a$ non vanishing then
there exist colour magnetic monopoles and the sources are not the
quarks but the gluons.

We thank to COLCIENCIAS for finantial support.
\section*{Appendix}
\subsection*{Appendix A: Generators of $SU(3)$ group}
The special unitary group in 3 dimensions $SU(3)$ has $3^2-1=8$
generators. These generators are labeled as $\lambda_1,
\lambda_2,...,\lambda_8$. The generators $\lambda_3$ and $\lambda_8$
are explicitly:
\begin{center}
 $
\lambda _3 = \left(
\begin{array}{ccc}
1 & 0 & 0\\
0 & -1 & 0\\
0 & 0 & 0
\end{array}
\right) $, \hspace{1cm}$ \lambda _8 = \frac{1}{\sqrt{3}}\left(
\begin{array}{ccc}
1 & 0 & 0\\
0 & 1 & 0\\
0 & 0 & -2
\end{array}
\right)
$.
\end{center}
\subsection*{Appendix B: Wave equations}
Using the SU(3)-Yang-Mills equations (\ref{yms}) and the
definition of the non-abelian gauge field tensor (\ref{ym1}), it
is possible to obtain:
\begin{eqnarray}
\partial ^\mu F^ a _{\mu \nu}&=&-gC^a _{bc}A^{b \mu} F^c _{\mu \nu},\notag\\
\partial ^\mu(\partial _\mu A^a _\nu -\partial _\nu A^a_\mu +gC^a_{bc}
A^b_{\mu} A^c_\nu)&=&-gC^a _{bc}A^{b \mu} (\partial _\mu A^c _\nu -
\partial _\nu A^c_\mu +gC^c_{mn}A^m_\mu A^n_\nu),\notag\\
\Delta A^a _\nu -\partial _\nu \partial ^\mu A^a_\mu +gC^a_{bc}
\partial ^\mu(A^b_{\mu} A^c_\nu)&=&-gC^a _{bc}A^{b\mu}
(\partial _\mu A^c _\nu -\partial _\nu A^c_\mu +gC^c_{mn}A^m_\mu A^n_\nu),\notag\\
\Delta A^a _\nu -\partial _\nu \partial ^\mu A^a_\mu +
gC^a_{bc}\partial ^\mu(A^b_{\mu})A^c_\nu&=&-gC^a _{bc}A^{b\mu}
(2\partial _\mu A^c _\nu -\partial _\nu A^c_\mu +gC^c_{mn}A^m_\mu
A^n_\nu).\hspace{1cm}
\end{eqnarray}
Using the gauge of Lorentz (fixing the gauge), we obtain a wave
equation given by
\begin{eqnarray}
\Delta A^a _\nu&=&gC^a _{bc}A^b _\mu (\partial _\nu
A^c_\mu-2\partial _\mu A^c _\nu-gC^c_{mn}A^m_\mu A^n_\nu).
\end{eqnarray}

\end{document}